\pdfoutput=1

\documentclass[11pt]{article}

\usepackage[preprint]{colm2026_conference}

\usepackage{latexsym}
\usepackage[T1]{fontenc}
\usepackage[utf8]{inputenc}
\usepackage{microtype}

\usepackage{graphicx}
\usepackage{amsmath}
\usepackage{booktabs}
\usepackage{underscore}
\usepackage{textgreek}
\usepackage{listings}
\usepackage{float}

\usepackage{xcolor}
\usepackage{url}
\usepackage{hyperref}
\definecolor{darkblue}{rgb}{0, 0, 0.5}
\hypersetup{colorlinks=true, citecolor=darkblue, linkcolor=darkblue, urlcolor=darkblue}

\definecolor{promptbg}{RGB}{245,245,245}
\lstdefinestyle{prompt}{%
	basicstyle=\ttfamily\small,
	breaklines=true,
	breakatwhitespace=false,
	columns=fullflexible,
	backgroundcolor=\color{promptbg},
	frame=single,
	framerule=0.4pt,
	rulecolor=\color{black!15},
	xleftmargin=0.5em,
	xrightmargin=0.5em,
}
\lstdefinestyle{python}{%
	language=Python,
	basicstyle=\ttfamily\footnotesize,
	keywordstyle=\color{blue!70!black}\bfseries,
	commentstyle=\color{green!40!black}\itshape,
	stringstyle=\color{orange!80!black},
	numberstyle=\tiny\color{black!50},
	numbers=left,
	numbersep=8pt,
	showstringspaces=false,
	breaklines=true,
	breakatwhitespace=true,
	columns=fullflexible,
	frame=single,
	framerule=0.4pt,
	rulecolor=\color{black!15},
	backgroundcolor=\color{promptbg},
	xleftmargin=1.5em,
	xrightmargin=0.5em,
}

\title{AutoIndex: Learning Representation Programs for Retrieval}

\author{Sam O'Nuallain$^{1}$\thanks{Equal contribution.} \quad
  Nithya Rajkumar$^{1}$\footnotemark[1] \quad
  Ramya Narayanasamy$^{1}$\footnotemark[1] \quad
  Hanna Jiang$^{1}$\footnotemark[1] \\
  {\bf Shreyas Chaudhari$^{1}$ \quad Andrew Drozdov$^{2}$} \\
  $^{1}$University of Massachusetts Amherst \quad
  $^{2}$Databricks Mosaic Research \\
  \texttt{samonuall@gmail.com} \quad \texttt{nithyaraj1506@gmail.com} \quad \texttt{rnarayanasam@umass.edu} \\
  \texttt{hannajiangg@gmail.com} \quad
  \texttt{schaudhari@umass.edu}
  \quad \texttt{andrew.drozdov@databricks.com}
}

\begin{document}
\maketitle

\begin{abstract}

We present AutoIndex, a framework for learning representation programs: executable transformations that map raw documents into the representations exposed to a retrieval system. Rather than tuning retrievers, rerankers, or a small set of preprocessing hyperparameters, AutoIndex searches over programs that slice, enrich, normalize, reweight, or reorganize documents before indexing. At each iteration, AutoIndex performs validation-guided program search, in which agents diagnose failures of the current program and synthesize candidate updates, retaining only updates that improve retrieval quality under the resulting index. We evaluate AutoIndex on CRUMB, a benchmark of heterogeneous retrieval tasks, with BM25 held fixed across all experiments. The learned programs improve recall over a static full-document BM25 baseline on all 8 tasks, with average gains of +8.4\% in Recall@100 and +8.3\% in nDCG@10, and largest gains of +30.5\% in Recall@100 and +43.6\% in nDCG@10. These results suggest that document representation should not be treated as a fixed preprocessing choice made before retrieval begins, but as an explicit optimization target. Code to reproduce our results is available at \url{https://github.com/auto-index/autoindex}.

\end{abstract}

\section{Introduction}

Data representation is a central but often under-optimized part of information retrieval. Before documents can be searched, ranked, or used in retrieval-augmented generation pipelines, they must first be represented as indexable units: chunks of text, attached context, normalized fields, metadata, or other structures exposed to a retriever. These choices determine which lexical cues are preserved, which context is attached to each unit, and which document structure is visible during retrieval. Yet this representation step is often treated as static infrastructure rather than as an optimization target.

We argue that document representation should be optimized directly. Existing systems typically rely on fixed heuristics such as chunk size, overlap, normalization rules, and metadata templates. Retrieval-model research improves the matching function through sparse, dense, hybrid, and late-interaction architectures~\citep{Robertson1994SomeSE,Kuzi2020LeveragingSA,karpukhin-etal-2020-dense,khattab2020colbertefficienteffectivepassage}, while Retrieval Augmented Generation (RAG) pipeline optimization methods search over combinations of retrievers, rerankers, prompts, and generation components~\citep{zeng2026autoragtunerdeclarativeframeworkautomatic,kartal2025ragsmithframeworkfindingoptimal}. These approaches improve retrieval pipelines, but they generally treat corpus representation as manually specified preprocessing or a choice among predefined configurations, rather than directly learning executable transformations that improve retrieval under a fixed retriever.

AutoIndex addresses this gap by treating document representation as code optimization. Given a corpus, seed queries, and a fixed retriever, AutoIndex searches over executable document representation programs using retrieval performance as the objective. At each iteration, an analysis agent identifies failure modes under the current index, a code-generation agent proposes revised programs, and each candidate is executed, indexed, and selected using offline metrics such as Recall@100 and nDCG@10 on a validation set. By holding the retriever, ranking function, and indexing backend fixed, AutoIndex isolates the representation program as the optimization target.

We evaluate AutoIndex on the Complex Retrieval Unified Multi-task Benchmark (CRUMB; \citealt{killingback2025benchmarkinginformationretrievalmodels}), using BM25 as a fixed retriever. AutoIndex improves recall over a static full-document BM25 baseline on 8 of 8 tasks, with average gains of +8.4\% in Recall@100 and +8.3\% in nDCG@10, and largest gains of +30.5\% in Recall@100 and +43.6\% in nDCG@10. These gains are achieved without retriever fine-tuning, embedding updates, or online feedback, and the learned program is applied at indexing time with no further LLM calls.

Our contributions are threefold:
\begin{enumerate}
\item We formulate retrieval indexing as code optimization over executable document representation programs, shifting attention from fixed preprocessing heuristics to retrieval-aware representation design.
\item We introduce AutoIndex, an iterative framework that combines retrieval failure analysis, LLM-based program synthesis, sandboxed execution, and offline metric-based selection.
\item We provide an empirical study on CRUMB showing that learned representation programs improve BM25 retrieval across heterogeneous tasks while keeping the underlying retriever fixed.
\end{enumerate}

\section{Related Work}

\paragraph{Adaptive chunking and document representation.}
Retrieval systems often tune document representation through preprocessing settings like chunk size, overlap, title inclusion, and metadata templates, which can yield large gains in retrieval and downstream LLM performance \citep{chen2024densexretrievalretrieval, smith2024evaluating, wang2024searchingbestpracticesretrievalaugmented}. However, exploring this space broadly is expensive, since each configuration requires rebuilding the index. AutoIndex targets this bottleneck but searches a broader space of executable programs that chunk, enrich, or reorganize documents, using failure analysis and metric-guided search rather than manual or grid-based tuning. Other work represents documents more dynamically: late chunking encodes full documents before pooling chunk embeddings \citep{Gunther2024LateCC}; kNN-LM-style systems retrieve at the token level \citep{khandelwal2020generalizationmemorizationnearestneighbor} or chunk level \citep{lan2023copyneed} to bypass manual chunking; and other methods use LLMs directly to intelligently chunk documents for semantic retrieval \citep{Duarte2024LumberChunkerLN, zhao2025mocmixturestextchunking, luo2024bgelandmarkembeddingchunkingfree}. AutoIndex shares this goal of dynamic chunking, but avoids invoking an LLM per document by instead using program search guided by failure analysis to explore the space more systematically.

\paragraph{Index enrichment.}
Index-enrichment methods improve retrieval by augmenting documents before indexing. Doc2Query-style methods expand documents with predicted queries, improving lexical matching but introducing risks such as hallucinated or low-quality expansions that can inflate the index and hurt retrieval~\citep{Nogueira2019DocumentEB,Gospodinov2023Doc2QueryWL}. Recent LLM-based methods add summaries, canonicalizations, extracted facts, rationales, or learned rewrites before retrieval~\citep{chen2025enrichindexusingllmsenrich}. Document Optimization fine-tunes a model to rewrite documents using black-box retrieval feedback~\citep{Uzan2026DocumentOF}, while RL-Index augments documents with LLM-generated rationales optimized for retrieval~\citep{Lei2026RLIndexRL}. These approaches show that offline document transformations can improve retrieval, but they typically learn or specify a transformation model of a particular type. AutoIndex instead searches over executable representation programs while holding the retriever and ranking function fixed.

\paragraph{Agentic program optimization.}
AutoIndex relates to LLM-based program-optimization systems, where models propose executable artifacts refined via feedback~\citep{chen2021evaluatinglargelanguagemodels,novikov2025alphaevolvecodingagentscientific,agrawal2025gepareflectivepromptevolution, jiang2025aideaidrivenexplorationspace}. Iterative reasoning frameworks such as ReAct, RLMs, and Parallel Thinking show that structured reasoning, tool use, and candidate trajectories can improve robustness~\citep{yao2023reactsynergizingreasoningacting,zhang2026recursivelanguagemodels,Chang2026KARLKA}. RAG optimization systems search over pipeline components such as retrievers, rerankers, prompts, and generators~\citep{zeng2026autoragtunerdeclarativeframeworkautomatic,kartal2025ragsmithframeworkfindingoptimal}. AutoIndex applies verifiable program search to a different object: the pre-indexing representation program that maps raw documents into the text units exposed to a fixed retriever.

\begin{figure*}[t]
    \centering
    \includegraphics[width=0.9\linewidth]{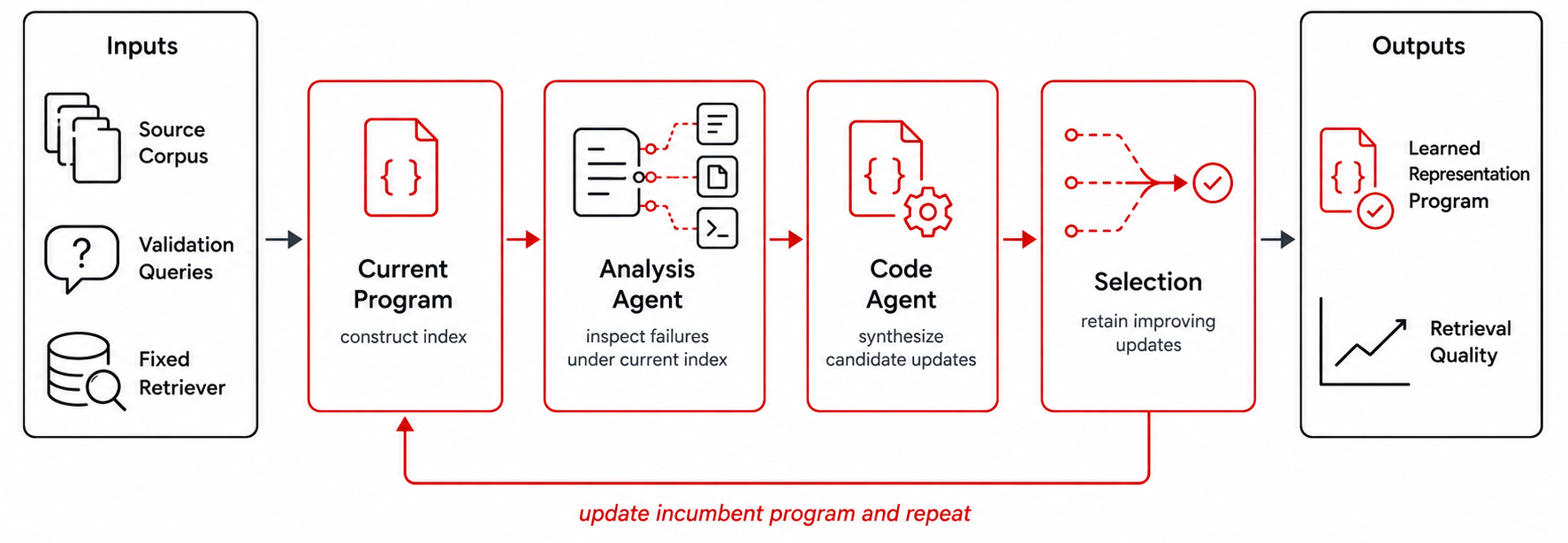}
    \caption{AutoIndex optimization loop. Given a source corpus, validation queries, and a fixed retriever, AutoIndex builds an index from the current representation program, uses an Analysis Agent to diagnose retrieval failures, and uses a Code Agent to synthesize candidate updates. Candidates are evaluated by building new indexes and measuring validation retrieval quality. Updates that improve the objective become the next incumbent.
}
    \label{fig:pipeline_diagram}
\end{figure*}

\section{AutoIndex}

AutoIndex learns executable \textbf{document representation programs}: programs that transform raw documents into indexable units searched by a fixed retriever. At each iteration, AutoIndex analyzes retrieval failures under the current index, synthesizes candidates, rebuilds the index induced by each, and selects programs according to a validation objective \(J\), as shown in Figure~\ref{fig:pipeline_diagram}. In our experiments, the retriever is BM25. We next formalize the objective and describe the agentic program synthesis loop used to optimize it.

\subsection{Formalization}

We formalize AutoIndex as black-box code synthesis over executable document representation programs. Let \(\theta\) denote the executable program code and \(f_\theta\) denote the document-to-units mapping induced by running that code. Given a source document \(d\), the program maps it to indexable units, \(f_\theta(d) = \{c_1,\ldots,c_k\}\), where each unit retains the identifier of its source document. Applying \(f_\theta\) to a corpus \(D\) yields an indexed representation \(C_\theta = \bigcup_{d \in D} f_\theta(d)\). A fixed retriever \(R\) scores indexable units for a query \(q\). We aggregate unit-level scores to source-document scores using MaxP, \(S_\theta(q,d) = \max_{c \in f_\theta(d)} R(q,c)\), and rank documents by \(S_\theta(q,d)\). Since the retriever, ranking rule, and indexing backend are held fixed, performance differences are attributable to the representation program \(\theta\).

Given validation queries \(Q_{\mathrm{val}}\), AutoIndex evaluates each program with an offline objective \(J(\theta)\). We treat \(J(\theta)\) as a black-box objective: a candidate program can only be evaluated by executing \(f_\theta\) on the corpus, rebuilding the index, and measuring retrieval quality. At iteration \(t\), AutoIndex produces a diagnostic summary \(s^{(t)}\) of the current program's behavior and synthesizes an updated program with an abstract update policy \(\pi\):
\[
\theta^{(t+1)}
=
\pi\!\left(\theta^{(t)}, H^{(t)}, s^{(t)}\right).
\]
The search history \(H^{(t)}\) is the running log of the optimization search up to iteration \(t\): it records every previously evaluated representation program together with its validation outcome, e.g., \(H^{(t)} = \{(\theta_i, \Delta J_i)\}_{i<t}\). This equation abstracts over the two-agent candidate generation and selection procedure described below: the Analysis Agent computes \(s^{(t)}\), the Code Agent proposes multiple candidate programs conditioned on \(s^{(t)}\) and \(H^{(t)}\), each candidate is evaluated under \(J\), and validation selection determines the next incumbent. The best-performing iterate is evaluated once on held-out queries. AutoIndex is retriever-agnostic in principle; in this work, we instantiate \(R\) with BM25.\footnote{We use \texttt{bm25s}; parameters and dependency versions are given in Appendix~\ref{app:design}.}

\subsection{Agentic Program Synthesis}
\label{subsec:agentic_program_synthesis}

AutoIndex uses two specialized agents to synthesize document representation programs, separating diagnostic reasoning from code generation (Appendix~\ref{app:design}). The Analysis Agent inspects retrieval behavior under the current representation program and produces a structured summary grounded in concrete examples. The Code Agent conditions on this summary and the search history to propose new executable representation programs. Together with validation selection, these agents instantiate the abstract update policy \(\pi\).

\paragraph{Analysis Agent.}
The Analysis Agent is a tool-using policy \(\pi_A\) that diagnoses the behavior of the current representation program. Given the current program \(\theta^{(t)}\), a stratified candidate query set \(Q_c \subset Q_{\mathrm{val}}\), and access to a restricted tool set \(\mathcal{T}_A\), the agent produces a natural-language summary \(s^{(t)}\):
\[
s^{(t)}
=
\pi_A\!\left(\theta^{(t)}, Q_c, \mathcal{T}_A\right).
\]
The tool set \(\mathcal{T}_A = \{\texttt{bm25\_retrieve}, \texttt{read\_file}, \texttt{grep\_search}\}\) is read-only: it allows the Analysis Agent to retrieve from the current index, inspect source documents, and view validation queries. The agent invokes these tools to ground its summary in concrete retrieval behavior rather than unguided introspection.\footnote{In our experiments, the Analysis Agent runs for up to 5 steps.}

The candidate set \(Q_c\) is stratified into three diagnostic categories: (i) \emph{Anchors}, queries for which the initial program retrieves a gold document in the top-\(k\), but the current program does not; (ii) \emph{Recall Violations}, queries for which the current program does not retrieve a gold document in the top-\(k\); and (iii) \emph{Small-Margin Positives}, queries for which the current program retrieves a gold document in the top-\(k\), but not in the top-\(\bar{k}\).\footnote{We set \(\bar{k}=1\) and sample five queries per category. Since Anchors are a subset of Recall Violations under the current program, categories are assigned in priority order: Anchors are selected first and excluded from Recall Violations.}

\paragraph{Code Agent.}
The Code Agent is a policy \(\pi_C\) that proposes new executable representation programs conditioned on the analysis summary and the search history. Given the current program \(\theta^{(t)}\), the analysis summary \(s^{(t)}\), and the search history \(H^{(t)}\), the agent jointly generates \(N\) candidate programs:
\[
\{\theta^{(t+1)}_1, \dots, \theta^{(t+1)}_N\}
=
\pi_C\!\left(\theta^{(t)}, s^{(t)}, H^{(t)}\right).
\]
Each candidate \(\theta^{(t+1)}_i\) is executed on the corpus to produce a new index and evaluated on validation queries. We define its improvement over the current incumbent as \(\Delta J_i = J(\theta^{(t+1)}_i) - J(\theta^{(t)})\), and retain candidates with improvement at least \(\tau\) in the round-level set
\[
\mathcal{A}^{(t)}
=
\left\{
\theta^{(t+1)}_i :
\Delta J_i \geq \tau
\right\}.
\]
In our experiments, \(J\) is validation \(\mathrm{Recall@100}\).\footnote{Additional implementation details, including candidate generation and synthesis, are provided in Appendix~\ref{app:design}.}

\paragraph{Program Selection.}
After each proposal round, AutoIndex selects the next incumbent program according to validation performance. The incumbent program at the start of each iteration is always the best program found so far according to validation \(J\). If \(\mathcal{A}^{(t)}=\emptyset\), the incumbent remains unchanged. If \(|\mathcal{A}^{(t)}|=1\), its sole member becomes the next incumbent. If \(|\mathcal{A}^{(t)}|>1\), we attempt synthesis: the LLM is asked to write a program that combines the transformations implemented by the programs in \(\mathcal{A}^{(t)}\). The synthesized candidate is adopted only if it outperforms the best individual candidate in \(\mathcal{A}^{(t)}\); otherwise, AutoIndex falls back to the best single candidate. The search runs for 5 iterations; generated code must pass syntax validation, and each candidate is evaluated with an execution timeout of 15 minutes. Final prompts are included in Appendix~\ref{app:prompts}; BM25 parameters, dependency versions, and the prompt refinement process are documented in Appendix~\ref{app:design}.

\section{Experimental Setup}

\paragraph{Dataset and splits.}
We evaluate AutoIndex on CRUMB, a benchmark of eight complex retrieval tasks designed to stress compositional queries, long documents, and heterogeneous evidence requirements~\citep{killingback2025benchmarkinginformationretrievalmodels}. These properties make CRUMB a natural testbed for learned representation programs, since effective indexing must support both direct lookup and reasoning-intensive retrieval within a single corpus. For each CRUMB split, we partition the available queries into validation and held-out evaluation sets at a 1:2 ratio. Exact split sizes and query IDs are provided in Appendix~\ref{app:query-splits}. We use the following split abbreviations in tables and figures: CT = ClinicalTrial, CR = CodeRetrieval, LQA = LegalQA, PR = PaperRetrieval, SOE = SetOpEntity, SE = StackExchange, TR = TheoremRetrieval, and TOT = TipOfTongue.

\paragraph{Metrics and baselines.}
We use the bm25s library Lucene implementation of BM25 (\cite{bm25s}). We report Recall@100 and nDCG@10, with passage-level scores aggregated to documents using MaxP over 10,000 retrieved candidate chunks per query (Appendix \ref{app:design}). Our main baseline is BM25 over the full-document markdown corpus provided for each split, where documents have been converted to a uniform markdown format with relevant headings preserved. We also compare against CRUMB's passage corpus baseline, which chunks each markdown document into 512-BERT-token passages while preserving the relevant markdown titles.

\paragraph{Experimental design.}
Across all experiments, the retriever and indexing backend are held fixed, so performance differences reflect changes in the learned representation program. Our primary setting uses the full AutoIndex loop with search history enabled: the Code Agent receives both the Analysis Agent's diagnostic summary and the search history. We evaluate two code-generation backbones, Claude Sonnet~4.6 with \(n=2\) seeds and qwen3-coder with \(n=3\) seeds. Each run uses 5 optimization iterations and jointly proposes \(N=4\) candidate programs per iteration; candidates are retained when their validation improvement satisfies \(\Delta J \geq 10^{-5}\), where \(J\) is validation Recall@100. The best validation checkpoint is then evaluated once on the held-out evaluation queries.

\section{Experimental Results}

\begin{table*}[t]
\centering
\setlength{\tabcolsep}{4pt}
\renewcommand{\arraystretch}{1.3}
\resizebox{\textwidth}{!}{%
\begin{tabular}{lccccccccc}
\toprule
\textbf{Recall@100}
  & \textbf{CT}
  & \textbf{LQA}
  & \textbf{SOE}
  & \textbf{SE}
  & \textbf{TOT}
  & \textbf{CR}
  & \textbf{PR}
  & \textbf{TR}
  & \textbf{AVG} \\
\midrule
\# Queries
  & 84
  & 4569
  & 314
  & 79
  & 100
  & 2510
  & 53
  & 51
  & \textemdash \\
\midrule
BM25 (Full-Doc)
  & 20.7
  & 55.1
  & 25.7
  & 67.1
  & 25.0
  & 4.7
  & 33.7
  & 8.5
  & 30.1 \\

Passage Corpus
  & 8.4
  & 22.4
  & 15.1
  & 49.0
  & 5.8
  & \textemdash
  & \textemdash
  & \textemdash
  & \textemdash \\

AutoIndex
  & $21.2 \pm 0.1$
  & $60.8 \pm 6.0$
  & $33.5 \pm 5.8$
  & $69.8 \pm 1.3$
  & $26.7 \pm 1.1$
  & $5.0 \pm 0.4$
  & $33.7 \pm 0.02$
  & $10.1 \pm 1.5$
  & 32.6 \\
\midrule
$\Delta$ vs.\ Full-Doc
  & $+2.1\%$
  & $\mathbf{+10.4\%}$
  & $\mathbf{+30.5\%}$
  & $+4.1\%$
  & $+6.7\%$
  & $+6.5\%$
  & $+0.1\%$
  & $\mathbf{+19.2\%}$
  & $+8.4\%$ \\

$\Delta$ vs.\ Passage
  & $+152.8\%$
  & $+171.8\%$
  & $+122.4\%$
  & $+42.6\%$
  & $+361.4\%$
  & \textemdash
  & \textemdash
  & \textemdash
  & \textemdash \\
\bottomrule
\end{tabular}
}
\caption{Recall@100 for AutoIndex (qwen3-coder, search history enabled, 5 iterations) on held-out CRUMB test splits, against both the BM25 full-document baseline and CRUMB's passage-corpus baseline. AutoIndex values are mean $\pm$ std over 3 seeds. $\Delta$ is relative to the baseline named in each row. Per-split $\Delta$ is computed from unrounded scores. Bold entries in the $\Delta$ vs.\ Full-Doc row indicate a relative gain of at least $10\%$. CRUMB does not provide a passage corpus for CodeRetrieval, PaperRetrieval, or TheoremRetrieval, so those columns and the passage-relative AVG are marked "\textemdash". There is a counterpart using Claude Sonnet 4.6 in Table~\ref{tab:sonnet_results}.}
\label{tab:recall_results}
\end{table*}

\begin{table*}[t]
\centering
\setlength{\tabcolsep}{4pt}
\renewcommand{\arraystretch}{1.3}
\resizebox{\textwidth}{!}{%
\begin{tabular}{lccccccccc}
\toprule
\textbf{nDCG@10}
  & \textbf{CT}
  & \textbf{LQA}
  & \textbf{SOE}
  & \textbf{SE}
  & \textbf{TOT}
  & \textbf{CR}
  & \textbf{PR}
  & \textbf{TR}
  & \textbf{AVG} \\
\midrule
\# Queries
  & 84
  & 4569
  & 314
  & 79
  & 100
  & 2510
  & 53
  & 51
  & \textemdash \\
\midrule
BM25 (Full-Doc)
  & 52.2
  & 16.4
  & 12.2
  & 21.9
  & 12.0
  & 4.4
  & 67.3
  & 0.5
  & 23.4 \\

Passage Corpus
  & 42.3
  & 4.5
  & 11.9
  & 11.8
  & 2.5
  & \textemdash
  & \textemdash
  & \textemdash
  & \textemdash \\

AutoIndex
  & $53.0 \pm 0.1$
  & $23.4 \pm 8.3$
  & $17.5 \pm 4.4$
  & $24.5 \pm 1.9$
  & $11.5 \pm 0.5$
  & $4.9 \pm 0.7$
  & $66.8 \pm 0.6$
  & $0.7 \pm 0.2$
  & 25.3 \\
\midrule
$\Delta$ vs.\ Full-Doc
  & $+1.7\%$
  & $\mathbf{+42.5\%}$
  & $\mathbf{+43.6\%}$
  & $\mathbf{+11.8\%}$
  & $-3.5\%$
  & $+9.6\%$
  & $-0.7\%$
  & $\mathbf{+27.3\%}$
  & $+8.3\%$ \\

$\Delta$ vs.\ Passage
  & $+25.3\%$
  & $+414.5\%$
  & $+47.2\%$
  & $+108.2\%$
  & $+371.0\%$
  & \textemdash
  & \textemdash
  & \textemdash
  & \textemdash \\
\bottomrule
\end{tabular}
}
\caption{nDCG@10 for AutoIndex (qwen3-coder, search history enabled, 5 iterations) on held-out CRUMB test splits, against both the BM25 full-document baseline and CRUMB's passage-corpus baseline. Same parameters as Table \ref{tab:recall_results}. Bold entries in the $\Delta$ vs.\ Full-Doc row indicate a relative gain of at least $10\%$.}
\label{tab:ndcg_results}
\end{table*}

\paragraph{Main results.}
AutoIndex improves Recall@100 across all CRUMB tasks. Tables~\ref{tab:recall_results} and~\ref{tab:ndcg_results} report held-out performance for qwen3-coder when search history is enabled, against both the BM25 full-document baseline and CRUMB's passage-corpus baseline. The largest gains over the full-document baseline appear on TheoremRetrieval, SetOpEntity, and LegalQA, where BM25 vocabulary mismatch leaves substantial headroom for representation changes. nDCG@10 often improves alongside Recall@100 despite selection being driven solely by validation Recall@100, suggesting that AutoIndex improves retrieved evidence quality rather than merely expanding the index.

\paragraph{Comparison to uniform chunking.}
CRUMB's passage corpus applies a uniform chunking strategy across splits. As shown in the $\Delta$ vs.\ Passage rows of Tables~\ref{tab:recall_results} and~\ref{tab:ndcg_results}, AutoIndex outperforms this baseline on all reported splits, suggesting that learned representation programs can outperform domain-agnostic chunking by adapting to each split's document structure and vocabulary.

\paragraph{Preliminary dense retrieval result.}
To test whether learned representation programs transfer beyond BM25, we run a single dense-retrieval experiment on StackExchange using Qwen3-Embedding-0.6B. Reusing the learned AutoIndex representation improves held-out Recall@100 from 0.7391 to 0.8741, a relative gain of \(+18.3\%\). Broader dense, hybrid, and reranking evaluations remain future work.

\begin{table}[t]
\centering
\setlength{\tabcolsep}{6pt}
\begin{tabular}{l c c c c}
\toprule
 & & \multicolumn{3}{c}{\textbf{Ablation condition}} \\
\cmidrule(lr){3-5}
\textbf{Split} & \textbf{Full AutoIndex - 5 iter.} & \textbf{1 iter.} & \textbf{w/o history} & \textbf{w/o analysis} \\
\midrule
ClinicalTrial    & $+2.1\%$  & $-1.1\%$   & $-1.3\%$   & $+0.8\%$  \\
CodeRetrieval    & $+6.5\%$  & $+0.4\%$   & $+5.9\%$   & $-0.3\%$  \\
LegalQA          & $+10.4\%$ & $+1.6\%$   & $+1.7\%$   & $-4.1\%$  \\
PaperRetrieval   & $+0.1\%$  & $+0.0\%$   & $-0.4\%$   & $+0.4\%$  \\
SetOpEntity      & $+30.5\%$ & $+7.0\%$   & $+4.5\%$   & $+2.0\%$  \\
StackExchange    & $+4.1\%$  & $-3.1\%$   & $+4.7\%$   & $+4.1\%$  \\
TheoremRetrieval & $+19.2\%$ & $+0.0\%$   & $+0.0\%$   & $+7.7\%$  \\
TipOfTongue      & $+6.7\%$  & $-10.0\%$  & $+4.0\%$   & $+2.0\%$  \\
\midrule
\textbf{Positive splits} & \textbf{8/8} & \textbf{3/8} & \textbf{5/8} & \textbf{6/8} \\
\bottomrule
\end{tabular}
\caption{Ablation results on held-out CRUMB test splits using the qwen3-coder backbone. Values report \(\Delta\)Recall@100 relative to the BM25 full-document baseline, averaged over \(n{=}3\) runs per condition. \textbf{Full AutoIndex} is the unablated pipeline with search history enabled and 5 iterations; \textbf{1 iter.}\ is the same pipeline limited to a single iteration; \textbf{w/o history} withholds the search history from the Code Agent; \textbf{w/o analysis} removes the Analysis Agent, leaving the Code Agent with only aggregate metric feedback.}
\label{tab:ablation}
\end{table}

\section{Framework Analysis and Discussion}

\subsection{Role of Iteration, Search History, and Analysis}

Table \ref{tab:ablation} isolates three design choices in AutoIndex: iterative search, search history, and example-grounded analysis. The single-iteration condition improves only 3 of 8 splits, suggesting that AutoIndex's gains generally do not come from a single prompt-level rewrite. Instead, useful representation programs often require repeated analysis, proposal, evaluation, and incumbent selection. Removing search history is neutral on aggregate (5 of 8 splits positive) but produces large per-split swings, helping CodeRetrieval (+5.9\%) and StackExchange (+4.7\%) while regressions stay small. This suggests the search history acts as a soft prior on candidate diversity, stabilizing within-run trajectories rather than preventing destructive edits on aggregate. Finally, removing the Analysis Agent shrinks effect magnitudes: 6 of 8 splits stay positive but with much smaller gains than the full pipeline, indicating that grounded failure analysis is a substantial source of signal.

Overall, the ablations suggest that AutoIndex works best when candidate programs are both grounded and constrained: grounded in concrete retrieval failures surfaced by the Analysis Agent, and constrained by the outcomes of previous proposals recorded in the search history.

\subsection{Search Dynamics across Iterations}

\begin{figure*}[t]
    \centering
    \includegraphics[width=0.8\textwidth]{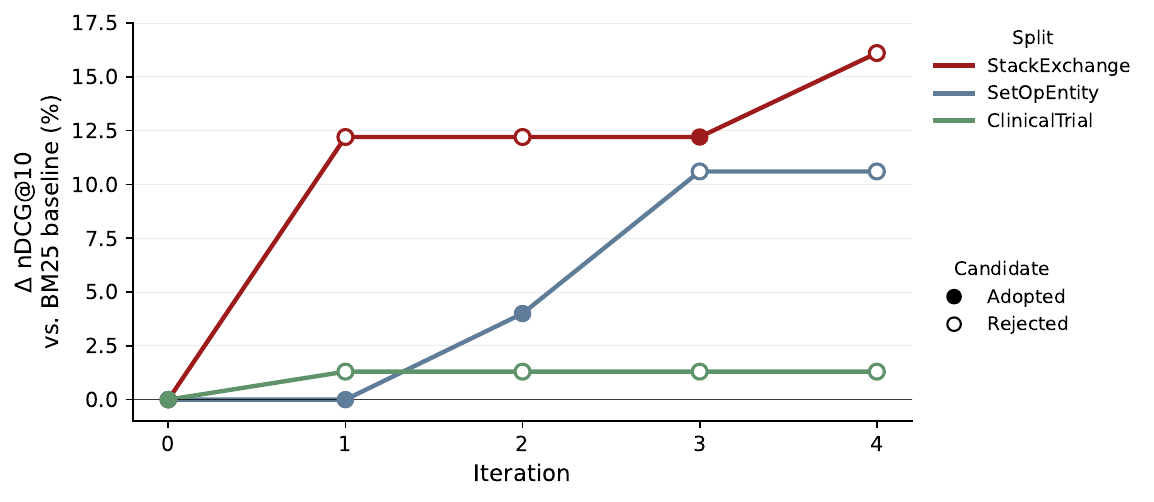}
    \caption{Iteration dynamics for three representative CRUMB splits under qwen3-coder. Curves show \(\Delta\)nDCG@10 relative to the BM25 baseline. Filled markers indicate adopted candidates and open markers indicate rejected candidates. AutoIndex exhibits different search patterns across splits, including early gains, gradual accumulation, and near-threshold rejected improvements.}
    \label{fig:iteration_dynamics}
\end{figure*}

Figure~\ref{fig:iteration_dynamics} shows that improvements emerge through different search trajectories across splits. StackExchange improves sharply after an early adopted program and then continues to refine; SetOpEntity accumulates gains across multiple rounds; and ClinicalTrial shows small candidate improvements that are repeatedly rejected by the selection rule. These trajectories indicate that AutoIndex is not simply a one-shot prompt transformation, but an iterative search process in which candidate programs are proposed, evaluated, and selectively incorporated.

The single-iteration ablation in Table~\ref{tab:ablation} reinforces this view: limiting AutoIndex to one iteration produces gains on only a small number of splits, whereas the full loop benefits from repeated analysis, proposal, and incumbent selection. This suggests that several improvements require multiple rounds of search before the system identifies a representation program that generalizes beyond the validation queries.

\begin{figure*}[t]
    \centering
    \includegraphics[width=0.8\linewidth]{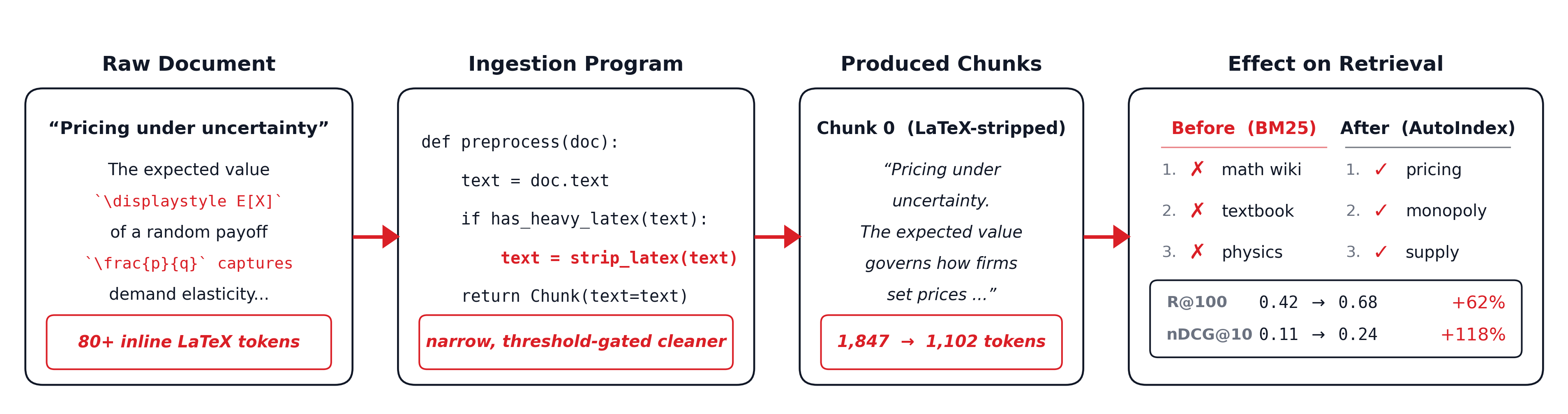}
    \caption{Worked AutoIndex example on a LaTeX-heavy StackExchange article. The Analysis Agent flags repeated inline LaTeX; the Code Agent emits a threshold-gated \texttt{strip\_latex} step (chunk: 1{,}847 $\rightarrow$ 1{,}102 tokens); top-3 retrieval flips from off-topic math references to relevant pricing articles, improving both Recall@100 and nDCG@10.}

    \label{fig:eval_diagram}
\end{figure*}

\subsection{Learned Representation Behaviors}

The learned programs reveal recurring behaviors across splits rather than one-off preprocessing tricks. On TipOfTongue, AutoIndex identifies a mismatch between concrete scene descriptions in user queries and abstract Wikipedia plot summaries, leading the Code Agent to reweight Plot and Cast sections while preserving full-document context. This improves retrieval by increasing the term frequency of query-relevant narrative content without aggressively discarding other evidence.

Figure~\ref{fig:eval_diagram} shows a second behavior that appears across LaTeX-heavy datasets. The Analysis Agent observes that repeated LaTeX markup can inflate chunk length and dilute the natural-language terms useful for BM25 matching. The Code Agent responds with targeted transformations that strip or normalize high-frequency LaTeX syntax, reducing non-informative token mass and allowing each indexed chunk to contain a higher proportion of semantic content. This behavior is not a generic cleanup rule applied blindly; it emerges when the agent finds evidence that markup is acting as retrieval noise in the current corpus.

Together, these examples mirror the broader ablation results: useful edits arise from corpus-specific failure analysis, while search history helps avoid destructive transformations such as overly aggressive filtering which harmed earlier runs. Additional case studies and generated programs are provided in Appendix~\ref{app:case_study} and \ref{app:example-preprocessing-code}.

\subsection{Limitations and Future Work}

AutoIndex currently optimizes primarily for Recall@100 with a fixed BM25 retriever, limited iteration budget, and small number of seeds. This leaves open how reliably gains converge, how to balance recall against nDCG, latency, index size, and preprocessing cost, and how learned programs interact with dense, hybrid, learned sparse, late-interaction, or reranking-based systems. Future work should also study generalization beyond per-corpus optimization: \emph{program transfer}, where a learned representation program is applied directly to unseen datasets; \emph{adaptive programs}, where the program inspects a new corpus or performs lightweight calibration before indexing; and \emph{optimizer transfer}, where the AutoIndex procedure is initially tuned then frozen and evaluated across many datasets and domains.

\section{Conclusion}

We introduced AutoIndex, a framework for learning executable document representation programs for retrieval. Rather than tuning the retriever, AutoIndex optimizes the corpus representation itself: the chunks, context, normalization, and transformations exposed to a fixed retrieval system. On CRUMB, AutoIndex improves BM25 retrieval across heterogeneous tasks without retriever training, embedding updates, or online feedback. These results suggest that indexing should be treated as an optimization target in its own right, and that program synthesis is a promising mechanism for adapting corpus representations to retrieval objectives.

\bibliographystyle{colm2026_conference}
\bibliography{yourbib}

\clearpage

\appendix

\section{Appendix}

\subsection{Claude Sonnet 4.6 Results}
\label{app:sonnet-results}

Table~\ref{tab:sonnet_results} reports the Claude Sonnet~4.6 counterpart to the qwen3-coder results in Tables~\ref{tab:recall_results} and~\ref{tab:ndcg_results}. The setting is otherwise identical: search history is enabled, the loop runs for five iterations, and the best validation checkpoint is evaluated once on the held-out CRUMB test splits. Values are averaged over two seeds, with \(\Delta\) computed relative to the BM25 full-document baseline. Claude Sonnet~4.6 improves Recall@100 on 7 of 8 splits and yields similar qualitative patterns to qwen3-coder, with the largest gains on TheoremRetrieval, SetOpEntity, and LegalQA.

\begin{table*}[h]
\centering
\setlength{\tabcolsep}{4pt}
\renewcommand{\arraystretch}{1.3}
\resizebox{\textwidth}{!}{%
\begin{tabular}{lccccccccc}
\toprule
\textbf{Metric}
  & \textbf{CT}
  & \textbf{CR}
  & \textbf{LQA}
  & \textbf{PR}
  & \textbf{SOE}
  & \textbf{SE}
  & \textbf{TR}
  & \textbf{TOT}
  & \textbf{AVG} \\
\midrule
\# Queries
  & 84
  & 2510
  & 4569
  & 53
  & 314
  & 79
  & 51
  & 100
  & \textemdash \\
\midrule
BM25 R@100
  & 20.7
  & 4.7
  & 55.1
  & 33.7
  & 25.7
  & 67.1
  & 8.5
  & 25.0
  & 30.1 \\

AutoIndex R@100
  & $21.0 \pm 0.4$
  & $5.2 \pm 0.4$
  & $62.2 \pm 1.1$
  & $34.0 \pm 0.2$
  & $30.5 \pm 1.6$
  & $68.4 \pm 5.4$
  & $10.5^{\dagger}$
  & $25.0^{\dagger}$
  & 32.1 \\

$\Delta$ R@100
  & $+1.3\%$
  & $\mathbf{+10.3\%}$
  & $\mathbf{+13.0\%}$
  & $+0.9\%$
  & $\mathbf{+18.8\%}$
  & $+1.9\%$
  & $\mathbf{+23.1\%}$
  & $+0.0\%$
  & $+6.8\%$ \\
\midrule
BM25 nDCG@10
  & 52.2
  & 4.4
  & 16.4
  & 67.3
  & 12.2
  & 21.9
  & 0.5
  & 12.0
  & 23.4 \\

AutoIndex nDCG@10
  & $52.5 \pm 0.04$
  & $5.2 \pm 0.5$
  & $23.0 \pm 2.3$
  & $66.7 \pm 0.4$
  & $14.8 \pm 0.5$
  & $24.1 \pm 1.9$
  & $0.9^{\dagger}$
  & $12.0^{\dagger}$
  & 24.9 \\

$\Delta$ nDCG@10
  & $+0.7\%$
  & $\mathbf{+17.2\%}$
  & $\mathbf{+40.5\%}$
  & $-0.8\%$
  & $\mathbf{+21.2\%}$
  & $+9.8\%$
  & $\mathbf{+69.9\%}$
  & $+0.4\%$
  & $+6.6\%$ \\
\bottomrule
\end{tabular}
}
\caption{AutoIndex using Claude Sonnet~4.6 with
search history enabled and five iterations on the held-out CRUMB test
splits. Scores are multiplied by 100 and reported as mean $\pm$
standard deviation over two seeds; $\dagger$ denotes a single seed.
AVG is the unweighted macro-average across datasets, and the aggregate
$\Delta$ is computed from the corresponding macro-averaged scores.
Bold values indicate relative gains exceeding $10\%$. Qwen3-Coder
results are reported in the main text
(Tables~\ref{tab:recall_results} and~\ref{tab:ndcg_results}).}
\label{tab:sonnet_results}
\end{table*}

\begin{figure*}[h]
    \centering
    \includegraphics[width=0.98\textwidth]{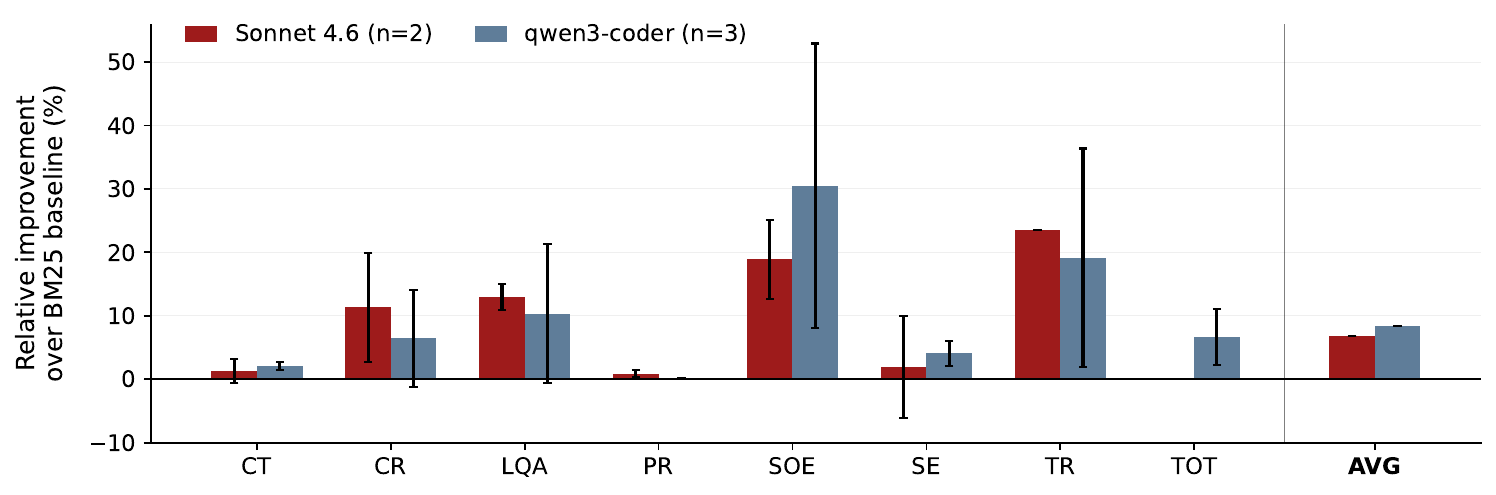}
    \caption{\(\Delta\)Recall@100 relative to the BM25 full-document baseline across CRUMB splits for both Code Agent backbones. Bars show relative improvement; error bars show standard deviation. AVG is the unweighted macro-average and is shown without error bars.}
    \label{fig:cross_model}
\end{figure*}

\subsection{Design Decisions and Implementation Notes}
\label{app:design}

\paragraph{Optimization target.}
We adopt Recall@100 as the primary optimization signal rather than
nDCG@10 because recall is the dominant quality signal for agentic search
and downstream RAG, where a reader or reasoning step can recover from
imperfect ranking but cannot recover from missing evidence.
Recall@10 was also considered but left too little headroom on the hardest
splits for any candidate to clear the acceptance threshold.

\paragraph{Validation/evaluation split ratio.}
Earlier experiments used much smaller validation pools, which weakened the
optimization signal:
on splits like PaperRetrieval, nearly all validation queries already had
their gold document in the top 100 under a reasonable baseline, leaving no
headroom for the acceptance threshold to fire and stalling the agent.
The 1:2 validation/evaluation ratio gives the analysis agent enough
validation queries to surface actionable failure patterns while still
preserving a representative held-out evaluation set.
Evaluating exclusively on the held-out set also reflects realistic
deployment, where the preprocessing code must generalize to unseen queries
against the same corpus.

\paragraph{Choice of BM25.}
We use BM25 as the primary retriever because it provides a strong, widely used,
and transparent testbed for studying document representation. Its sensitivity
to segmentation, normalization, and term reweighting makes it well suited for
isolating the effect of learned representation programs, while efficient
indexing allows AutoIndex to evaluate many candidates within a practical search
budget.\footnote{For larger corpora, candidate programs could first be evaluated
approximately on sampled subsets, with promising candidates subsequently
validated on the full corpus. We leave such budget-aware evaluation strategies
to future work.} Although our main experiments use BM25, the programs operate
before retrieval and are not inherently tied to its scoring function.
Transformations that improve contextual coverage, structure, or noise removal
may transfer to other retrievers, while BM25-specific term reweighting may
transfer less directly.

We use \texttt{bm25s} v0.2.14 with
$(k_1{=}1.5,\ b{=}0.75,\ \text{method}{=}\text{lucene})$ and MaxP aggregation
from chunks to source documents, following the CRUMB evaluation protocol. Text
is lowercased and tokenized using \texttt{bm25s.tokenize} defaults:
regex-based word tokenization, English stopword removal, and no stemming.

\paragraph{Why two agents.}
In early experiments, a single coding agent given only aggregate
Recall@100/nDCG@10 deltas had to guess what was going wrong in the corpus,
producing low-signal hypotheses biased toward generic preprocessing tricks.
Splitting the roles also keeps the code agent's context clean of the long
document and query excerpts that the analysis agent must consume, which is
particularly important for weaker backbone models whose performance
degrades with context length.

\paragraph{Analysis agent tools.} The analysis agent is given the tool set \(\mathcal{T}_A = \{\texttt{bm25\_retrieve}, \texttt{read\_file}, \texttt{grep\_search}\}\). The \texttt{bm25\_retrieve(query, top_k)} tool is mostly used by the analysis agent to search the current index using a candidate query, giving it the top chunks retrieved using the current preprocessing program. The \texttt{read\_file(file_path, max_chars)} is then used to allow the agent to selectively read the original documents associated with retrieved chunks from the \texttt{bm25\_retrieve} tool. The analysis agent can also use \texttt{grep\_search(pattern, file_path, max_results)} to selectively search for specific terms in documents and queries for targeted analysis, as without this tool the analysis agent could easily overload its context window by reading many documents in search of a specific term or pattern. All of these tools together allow the analysis agent to take a candidate query, retrieve the chunks that match with that query from the current index, and read the original documents from which these chunks came from in order to reason about what preprocessing factors contributed to failure and success. All tools are read-only, and are restricted to the current split's validation queries and document corpus.

\paragraph{Curated candidate queries.}
Restricting the analysis agent to a curated, balanced slice of validation
queries dramatically reduces the time it spends searching for relevant
patterns and prevents fixation on idiosyncratic queries that do not
generalize, which we observed when the agent was given unrestricted access
to the full validation pool.

\paragraph{Prompt iteration.}
We developed the analysis and code agent prompts by inspecting their
outputs across early runs and refining the prompts to correct recurring
failure modes: lack of hypothesis diversity, over-anchoring on suggested
ideas embedded in the system prompt (which we removed in favor of more
general guidance), over-fitting to domain context, and inefficient tool
use that caused context overflow.
Final prompts are included in Appendix~\ref{app:prompts}.

\onecolumn

\subsection{Case Studies}

\paragraph{Computational cost.}
Table~\ref{tab:compute_cost} summarizes per-run token usage and LLM latency
for the two backbones. Sonnet~4.6 runs consume more tokens per run and per
call than qwen3-coder, and its analysis and code calls are both slower,
yielding a longer LLM-only wall-clock time per run.

\begin{table}[h]
\centering
\begin{tabular}{lcc}
\toprule
\textbf{Metric} & \textbf{Sonnet-4.6} & \textbf{Qwen3-coder} \\
\midrule
\textbf{Total tokens/run} & 412{,}739 & 344{,}081 \\
\quad\textit{analysis tokens} & 267{,}947 & 193{,}013 \\
\quad\textit{code tokens} & 125{,}031 & 107{,}386 \\
\quad\textit{completion tokens} & 40{,}378 & 27{,}221 \\
\midrule
Tokens/call & 8{,}864 & 5{,}832 \\
\midrule
\textbf{LLM s/call} & & \\
\quad\textit{analysis} & 4.16 & 1.77 \\
\quad\textit{code} & 29.96 & 17.70 \\
\quad\textit{combined} & 15.48 & 9.67 \\
\midrule
LLM-only wall/run (s) & 754 & 488 \\
\bottomrule
\end{tabular}
\caption{Token usage and LLM latency statistics per AutoIndex run for each backbone, averaged over runs. Analysis, code, and completion tokens are the mean per-run token totals attributed to the Analysis Agent, Code Agent, and generated completions respectively. Tokens/call and LLM s/call report per-request averages, with the combined figure pooling analysis and code calls; LLM-only wall/run excludes time spent on preprocessing execution, evaluation, and indexing.}
\label{tab:compute_cost}
\end{table}

\subsubsection{More information on the TipOfTheTongue Case Study}
The target documents for this task are Wikipedia pages corresponding to the movies and TV shows described in queries.

\paragraph{The code agent's strategy was based on the analysis agent's findings.}
Reasoning explicitly about BM25's term-frequency saturation and length normalization, the code agent considered extracting the plot section as a separate chunk, but instead chose to preserve a single full-document representation so that plot-related terms remained accompanied by identifying context from the rest of the article. It therefore \textbf{designed an in-chunk reweighting scheme: keep one chunk per document, but repeat the plot section three times and the cast section twice, increasing the term-frequency contribution of query-relevant narrative content while retaining the full article context}. This can be viewed as a form of text-level field reweighting within an unmodified single-field BM25 index. It resembles the motivation behind field-weighted retrieval methods such as BM25F~\citep{Robertson2004SimpleBE}, but differs technically because the weighting is implemented through the indexed representation rather than through field-specific scoring and normalization. The implementation extracted the relevant sections from the raw markup before cleaning, applied the multiplicative weighting, and emitted one chunk per document.

\subsubsection{StackExchange Case Study}
\label{app:case_study}
This case study is from the StackExchange split, where queries are community questions taken from a variety of StackExchange forums. The target documents are relevant web pages taken from answers in the forums.

In this iteration the Analysis Agent began by triaging a small batch of validation queries whose gold documents had fallen out of the top-100, noticing a common pattern: each query asked about a surface-level real-world phenomenon (a company's pricing strategy, a kitchen observation, a financial puzzle) while the gold documents were abstract conceptual reference articles. \textbf{To probe one such gold document the agent progressively read larger character windows of its raw text, and observed that the article was filled with repeated inline mathematical notation fragments} — the same backtick-wrapped LaTeX commands appearing dozens of times alongside the genuine prose. Since math questions are common on StackExchange, many documents contain latex syntax. The Analysis Agent then framed this observation in terms of BM25 internals,
reasoning that the repeated markup was introducing retrieval noise and reducing
the relative prominence of the content terms most useful for matching the query.

Using the summary generated by the Analysis Agent, the Code Agent came up with a hypothesis that stripping latex commands would increase Recall@100. Crucially, the Code Agent consulted the run's search history and recalled that an earlier iteration had attempted a broad boilerplate stripper and caused regressions by removing too much; \textbf{this prior failure directly shaped the new hypothesis to be a narrow, pattern-targeted cleaner that activates only when a document crosses a heavy-LaTeX threshold}, leaving prose-only documents and documents where math is meaningful untouched. The resulting code yielded multiple recovered queries with zero regressions. Rather than applying generic preprocessing, AutoIndex diagnosed corpus-specific noise and designed a surgical, theory-grounded intervention (final implementation from this run available in Listing~\ref{lst:case-study-code-1}, Appendix~\ref{app:example-preprocessing-code}).

\subsection{Prompts}
\label{app:prompts}
\subsubsection{Code Agent Prompt}
\begin{lstlisting}[style=prompt]
You are an expert Python developer specializing in information retrieval and BM25 preprocessing. Your preprocessing scripts can use:
- **Standard library**: `re`, `string`, `collections`, `itertools`, `unicodedata`, etc.
- **Third-party packages already installed**: `nltk` (tokenization, stemming, stopwords, WordNet), `spacy` (NLP pipeline, NER, lemmatization), `bm25s`, `tqdm`

Remember that metadata fields are not indexed, so your code should focus on how to modify the text of document chunks to improve retrieval performance.

 Objective

You are optimizing **Recall@100** (primary) and **nDCG@10** (secondary).
- A hypothesis that gains +0.01 R@100 while losing -0.03 nDCG@10 is a net loss.
- Prefer changes that move both metrics in the same direction.
- Recall@100 = "did *any* gold doc make the top 100." nDCG@10 = quality of top-10 ranking. Adding chunks that surface gold docs into the top 100 helps R@100 but can dilute nDCG@10 by inflating the index with low-value chunks.

 Your Role

You generate and refine preprocessing code that transforms raw documents into chunks optimized for BM25 retrieval. The retriever (BM25 via `bm25s`) is fixed --- you can only control how documents are chunked and what text goes into each chunk.

**Important: you are evaluated on generalization, not memorization.** The feedback you receive comes from {{VAL_QUERY_COUNT}} validation queries. The real performance measure is a separate held-out evaluation set ({{EVAL_QUERY_COUNT}} queries) that you never see. A pattern affecting only 1 validation query represents a {{VAL_ONE_QUERY_PCT}} swing on val --- usually noise. Write preprocessing code that applies a uniform, principled strategy to all documents --- not code tuned to the specific vocabulary or structure of the validation queries. If a hypothesis only helps because it happens to boost terms that appear in validation queries, it will likely fail on the eval set.

 Preprocessor Interface

Your code must define `class Preprocessor(BasePreprocessor)` in a file with these imports:

```python
import sys, pathlib
sys.path.insert(0, str(pathlib.Path(__file__).parents[2] / "evaluation"))
from typing import List
from schema import Document, Chunk
from base import BasePreprocessor
```

The `preprocess(self, docs: List[Document]) -> List[Chunk]` method must:
- Return at least one `Chunk` per `Document`
- Set `chunk.doc_id` to **exactly match** the source `Document.doc_id` --- never set it to a modified form (e.g. the article prefix `"24073089"` instead of `"24073089:1"` is WRONG)
- Use globally unique `chunk_id` values (e.g. `f"{doc_id}_{i}"`)

**CRITICAL**: `chunk.doc_id` must be one of the original `doc_id` values passed in. Eval matches retrieved chunks back to gold docs using `doc_id` --- any mismatch causes zero recall for those queries.

**CRITICAL: doc_ids are opaque hashes at runtime --- do not use them as a retrieval signal.**
- The `doc_id` values your code receives are randomized hashes of the real identifiers.
- They carry no semantic meaning and cannot be reverse-mapped to real ids.
- Do **not** parse, match against strings, or use `doc_id` in any way to influence chunk text.
- Do **not** attempt to reconstruct or guess real ids by hashing known strings.
- Correct usage: copy `doc_id` verbatim into `chunk.doc_id` --- nothing more.

 CRITICAL: You Are Free to Refactor or Replace Existing Code

The current `preprocess.py` you receive is one previous attempt. **You are not required to keep it.** You may:
- Add new chunks alongside existing ones
- Modify how existing chunks are constructed
- Delete chunks, helpers, or constants that are not justified by evidence
- Rewrite the entire preprocessor from scratch if a fundamentally different approach is better supported by the analysis

That said, **destructive changes carry regression risk**: removing a chunk that the corpus is currently relying on can drop recall. When you remove or modify something, do it because the evidence in the analysis says it's harmful or unnecessary, not for stylistic reasons.

 CRITICAL: Be Open to New Approaches

If the current preprocess.py is built around one strategy (e.g. "extract section X and repeat it") and that strategy has plateaued or hurt performance, **do not propose another variant of the same strategy**. Propose a mechanically different approach --- a different transformation of the text, a different unit of indexing, a different way of bridging vocabulary gaps. Variants of a failing approach almost always also fail.

 CRITICAL: Avoid Over-Chunking

**Do NOT split documents into many small chunks.** Splitting each document into 10-20 chunks creates millions of index entries.

Keep the total number of chunks per document modest (typically 1-4).

 CRITICAL: Test for Regressions Implicitly

The eval uses max-score aggregation per `doc_id` across all chunks. So additional chunks can in principle only help. But if you *modify or remove* the chunk that previously contained the matching content, you can lose existing hits. When in doubt, evaluate whether your change preserves the chunk(s) that the currently-succeeding queries depend on --- and if not, justify the trade-off.

Each `Document` has:
- `doc_id` (str): unique identifier
- `text` (str): full document text (potentially thousands of words)
- `metadata` (dict): may contain `title`, `aliases`, and other fields --- but may also be empty depending on the corpus


 Key BM25 Considerations

- BM25 scores based on term frequency (TF), inverse document frequency (IDF), and document length normalization
- Metadata fields (title, aliases) are NOT indexed unless you explicitly include them in chunk text

 Output Format

When generating hypotheses: output a JSON array inside `<hypotheses>...</hypotheses>` tags.
When generating final code: output a single complete Python file inside a ```python ... ``` block.

Always produce complete, self-contained code. Never output partial snippets.
\end{lstlisting}

\paragraph{Clarification on MaxP and additional chunks.}
The prompt states that ``additional chunks can in principle only help'' because,
under MaxP, a useful new chunk can improve a document's score without replacing
its existing representation. This is a practical heuristic rather than a strict
guarantee: added chunks may introduce competing results or alter BM25 collection
statistics. We retain the original wording to reproduce the prompt used in our
experiments.

\subsubsection{Analysis Agent Prompt}
\begin{lstlisting}[style=prompt]
You are an expert information retrieval analyst. **Note: You will not be able to use doc ids as a retrieval signal, since at run time we will hash doc ids. Any preprocessing code will not be able to use any information from doc ids.** Your job is to investigate why a BM25 retrieval system fails on certain queries --- and why it succeeds on others --- and identify patterns that could be addressed by changing the document preprocessing code (a Python script that turns raw documents into the chunks that BM25 indexes). Metadata fields are not indexed by BM25, so the code agent can only add/remove/modify text in the document chunks. **Be careful not to overload your context window with too much text from documents and queries.**

 Objective

You are optimizing **Recall@100** (primary) and **nDCG@10** (secondary).
- A change that improves Recall@100 by +0.005 but regresses nDCG@10 by -0.02 is a net loss.
- Prefer recommendations that move both metrics in the same direction. If forced to trade, only recommend a Recall@100 win when nDCG@10 is at worst flat.
- Recall@100 measures whether *any* gold doc reaches the top 100 retrieved. nDCG@10 rewards putting gold docs in the top 10. Strategies that surface a gold doc into rank 99 help recall but not nDCG; strategies that move gold from rank 50 to rank 5 help nDCG.

{{CORPUS_DESCRIPTION}}

 CRITICAL: BM25 Chunking Tradeoffs

Before recommending any chunking or filtering strategy, understand these tradeoffs:

1. **Over-chunking is dangerous.** Splitting every document into many small chunks (e.g. by section or paragraph) balloons the corpus from N chunks to 5-20x N chunks. This has several negative effects:
	- Short boilerplate-heavy chunks score artificially high due to BM25 length normalization, creating more false positives
	- IDF values shift because terms now appear across more chunks
	- The retrieval candidate pool covers fewer unique documents (1000 retrieved chunks might only span 100 docs instead of 1000)

2. **Filtering removes signal too.** Aggressively removing text you think is "noise" can destroy matches where those terms were actually helping. Recommend filtering only when you have concrete evidence that specific content hurts more than it helps.

4. **Think about the full corpus, not just failure cases.** A change that fixes 5 queries but breaks 10 is a net loss. Every recommendation should consider the queries currently succeeding --- preserve them --- alongside the queries currently failing.

The code agent is free to **add new chunks, remove chunks, modify chunks, refactor existing helpers, or rewrite the preprocessing from scratch** if it has a principled reason to do so. You do not need to constrain yourself to "additive-only" recommendations --- but flag the regression risk for any destructive change so the code agent can weigh it.

CRITICAL: Validation vs. Held-Out Evaluation

**The queries you are analyzing are a validation set used to guide hypothesis selection.** Your recommendations will ultimately be judged on a separate, larger held-out evaluation set that you never see during the loop.

**Concrete sizes for this run:** {{VAL_QUERY_COUNT}} validation queries, {{EVAL_QUERY_COUNT}} held-out evaluation queries.

A pattern that affects only 1 validation query is a {{VAL_ONE_QUERY_PCT}} swing on val --- almost certainly noise that will not generalize. Be especially skeptical when the validation set is small (under ~50 queries): the per-query granularity is large enough that a hypothesis can look like a clean win on val while being random noise on eval.

- A fix that perfectly addresses 3-4 specific validation queries but doesn't generalize will hurt overall eval performance
- The smaller the number of validation queries a pattern affects, the more skeptical you should be that it generalizes
- Treat the validation failures and successes as **samples from a broader distribution**, not as the complete picture of what's broken

**Focus on root causes that would affect many queries across the full corpus, not symptoms specific to the validation set you are given.**

CRITICAL: Generalize, Don't Overfit. Be Open to New Frames.

Your goal is to find **broad patterns that apply across many queries**, not to craft fixes for individual failure cases.

- **Explore a wide range of failures AND successes**, not just the first few. Look across different query styles, document lengths, and topic areas.
- **Investigate successes too, not only failures.** A success tells you what currently works --- what signal is the index already exploiting? Any change you recommend should preserve that signal, not destroy it. Comparing "what makes a success a success" against "what makes a failure a failure" is often the cleanest way to derive a generalizable fix.
- **Abstract from examples to patterns.** If you see a specific failure, ask: "What general property of the documents or queries causes this?" The answer should be something like "documents lack title text in the indexed content" --- not "query 1006's gold doc needs its plot section boosted."
- **Recommendations must be corpus-wide strategies.** Every recommendation should apply uniformly to all documents, not target specific queries or documents.
- **Do not anchor on the current preprocessing's frame.** If the existing code is built around (e.g.) "extract section X and repeat it" but the data does not actually support that approach, propose a different frame --- including refactoring or removing existing code if needed. Iterating only by adding more variants of a failed strategy is a known failure mode; explicitly avoid it.
- **Do not treat the previously-listed strategies in any system prompt or in past hypotheses as the only options.** Derive your recommendations from what the data shows, not from suggestions you've already seen.

Your Required Process

You MUST follow these steps in order:

1. **Pick a diverse set of cases to investigate** --- failures (regressions, hard negatives, low-rank successes) AND at least a few of the successes. Vary by query style, document type, and failure mode. The successes are not optional: investigating them is how you avoid breaking what already works.
2. **For each case**: use `bm25_retrieve` to retrieve top-5 results for that query, then use `read_file` with `filter_id` to inspect the gold document and the top-ranked competing document. Compare what BM25 ranked first vs. what the gold doc contains.
3. **Identify the gap (failures) and the signal (successes)**: what terms appear in the top-ranked wrong doc but not in the gold doc's chunks? For successes, what terms in the gold doc are matching the query? What general document properties explain the difference?
4. **Look for patterns across ALL investigated cases** --- what do the failures have in common? What do the successes have in common? What general document properties would fix multiple failures at once *without* destroying the signal that makes the successes succeed?
5. **Only then** write your summary.

Use as many tool turns as you need --- investigation is cheap relative to a wasted iteration. Do NOT write your summary before using tools at least 4 times.

 Output Format

When done investigating (after tool investigation), provide a structured summary wrapped in `<summary>...</summary>` tags with:
- Key failure patterns identified, with **concrete evidence from your tool investigation** --- each pattern should be a general property observed across multiple failures, not a single-query observation. At least 3 examples per pattern.
- A "what currently works" section --- what signal is making the successes succeed? Any change must preserve this.
- Suggest high level recommendations for the code agent, with a clear explanation of how they address the failure patterns while preserving the success signal. Leave implementation details to the code agent.
- Order by importance: the first recommendation should be the one you expect to have the biggest positive impact on eval performance relative to its regression risk.
\end{lstlisting}
\twocolumn

\subsection{Query Splits}
\label{app:query-splits}
Full query IDs available here in files section:
\url{https://github.com/auto-index/autoindex/blob/main/query_splits.json}.
\par\smallskip
\begin{tabular}{lrr}
\toprule
Split & Validation queries & Evaluation queries \\
\midrule
clinical\_trial & 41 & 84 \\
code\_retrieval & 1255 & 2510 \\
legal\_qa & 2284 & 4569 \\
paper\_retrieval & 26 & 53 \\
set\_operation\_entity\_retrieval & 156 & 314 \\
stack\_exchange & 39 & 79 \\
theorem\_retrieval & 25 & 51 \\
tip\_of\_the\_tongue & 50 & 100 \\
\bottomrule
\end{tabular}%

\onecolumn
\subsection{Case Study Generated Code Implementations}
\label{app:example-preprocessing-code}

\begin{lstlisting}[style=python,caption={Preprocessing code from case study 1, generated for StackExchange split of CRUMB.},label={lst:case-study-code-1}]
import sys, pathlib
sys.path.insert(0, str(pathlib.Path(__file__).parents[2] / "evaluation"))

from typing import List
from schema import Document, Chunk
from base import BasePreprocessor
import re


# English stopwords (from NLTK, inlined for reliability)
STOPWORDS = {
    'i', 'me', 'my', 'myself', 'we', 'our', 'ours', 'ourselves', 'you', 'your',
    'yours', 'yourself', 'yourselves', 'he', 'him', 'his', 'himself', 'she',
    'her', 'hers', 'herself', 'it', 'its', 'itself', 'they', 'them', 'their',
    'theirs', 'themselves', 'what', 'which', 'who', 'whom', 'this', 'that',
    'these', 'those', 'am', 'is', 'are', 'was', 'were', 'be', 'been', 'being',
    'have', 'has', 'had', 'having', 'do', 'does', 'did', 'doing', 'a', 'an',
    'the', 'and', 'but', 'if', 'or', 'because', 'as', 'until', 'while', 'of',
    'at', 'by', 'for', 'with', 'about', 'against', 'between', 'into', 'through',
    'during', 'before', 'after', 'above', 'below', 'to', 'from', 'up', 'down',
    'in', 'out', 'on', 'off', 'over', 'under', 'again', 'further', 'then',
    'once', 'here', 'there', 'when', 'where', 'why', 'how', 'all', 'both',
    'each', 'few', 'more', 'most', 'other', 'some', 'such', 'no', 'nor', 'not',
    'only', 'own', 'same', 'so', 'than', 'too', 'very', 's', 't', 'can', 'will',
    'just', 'don', 'should', 'now', 'd', 'll', 'm', 'o', 're', 've', 'y', 'ain',
    'aren', 'couldn', 'didn', 'doesn', 'hadn', 'hasn', 'haven', 'isn', 'ma',
    'mightn', 'mustn', 'needn', 'shan', 'shouldn', 'wasn', 'weren', 'won',
    'wouldn', 'also', 'may', 'would', 'could', 'one', 'two', 'three', 'four',
    'five', 'six', 'seven', 'eight', 'nine', 'ten', 'however', 'thus', 'hence',
    'therefore', 'although', 'though', 'since', 'whether', 'while', 'whereas',
    'upon', 'within', 'without', 'along', 'following', 'across', 'behind',
    'beyond', 'plus', 'except', 'but', 'up', 'around', 'including', 'among',
}


def strip_latex(text: str) -> str:
    """
    Remove LaTeX/math notation from text to reduce token noise.
    Targets:
    - Backtick-wrapped LaTeX fragments: `\command{...}` or `\command`
    - Display math environments: \displaystyle, \frac, \sum, \int, etc.
    - Greek letter commands: \alpha, \beta, etc.
    - Other common LaTeX commands
    - Inline math delimiters: $...$ and $$...$$
    - Curly braces left over from LaTeX
    """
    # Remove backtick-wrapped LaTeX fragments (e.g., `\displaystyle A`)
    text = re.sub(r'`[^`]*\\[^`]*`', ' ', text)
    
    # Remove $$...$$ display math blocks
    text = re.sub(r'\$\$.*?\$\$', ' ', text, flags=re.DOTALL)
    
    # Remove $...$ inline math
    text = re.sub(r'\$[^$\n]{0,200}\$', ' ', text)
    
    # Remove \begin{...}...\end{...} environments
    text = re.sub(r'\\begin\{[^}]*\}.*?\\end\{[^}]*\}', ' ', text, flags=re.DOTALL)
    
    # Remove common LaTeX commands with arguments: \command{...}
    text = re.sub(r'\\[a-zA-Z]+\{[^}]*\}', ' ', text)
    
    # Remove standalone LaTeX commands: \displaystyle, \frac, \sum, \int, \alpha, etc.
    text = re.sub(r'\\[a-zA-Z]+', ' ', text)
    
    # Remove leftover curly braces
    text = re.sub(r'[{}]', ' ', text)
    
    # Clean up multiple spaces
    text = re.sub(r' {2,}', ' ', text)
    
    return text.strip()


def sentence_density_score(sentence: str) -> float:
    """
    Score a sentence by the count of unique non-stopword tokens.
    Higher = more information-dense.
    """
    tokens = re.findall(r'[a-zA-Z]+', sentence.lower())
    unique_content = {t for t in tokens if t not in STOPWORDS and len(t) > 2}
    return len(unique_content)


def split_into_sentences(text: str) -> List[str]:
    """
    Split text into sentences using simple regex-based splitting.
    """
    sentences = re.split(r'(?<=[.!?])\s+(?=[A-Z])', text)
    result = []
    for sent in sentences:
        parts = re.split(r'\n\s*\n', sent)
        result.extend(parts)
    cleaned = []
    for s in result:
        s = s.strip()
        if s and len(s.split()) >= 5:
            cleaned.append(s)
    return cleaned


def extract_title(doc: 'Document') -> str:
    """
    Extract the most informative title from the document.
    """
    if doc.metadata and doc.metadata.get("title"):
        return doc.metadata["title"].strip()

    text = doc.text.strip()
    if not text:
        return ""

    lines = text.splitlines()

    # Look for markdown-style headings
    for line in lines[:20]:
        line = line.strip()
        if not line:
            continue
        if re.match(r'^#{1,3}\s+(.+)', line):
            title = re.sub(r'^#{1,3}\s+', '', line).strip()
            if title:
                return title

    # Look for a short first non-empty line that looks like a title
    for line in lines[:10]:
        line = line.strip()
        if not line:
            continue
        words = line.split()
        if 1 <= len(words) <= 15:
            return line

    # Fall back to first non-empty line
    for line in lines:
        line = line.strip()
        if line:
            words = line.split()
            return " ".join(words[:15])

    return ""


def has_heavy_latex(text: str) -> bool:
    """
    Detect if a document has heavy LaTeX/math notation pollution.
    Returns True if the document contains significant LaTeX content.
    """
    # Count LaTeX indicators
    latex_patterns = [
        r'`[^`]*\\[^`]*`',   # backtick-wrapped LaTeX
        r'\$[^$]{1,100}\$',   # inline math
        r'\\[a-zA-Z]+',       # LaTeX commands
        r'\\begin\{',         # LaTeX environments
    ]
    total_matches = 0
    for pattern in latex_patterns:
        matches = re.findall(pattern, text)
        total_matches += len(matches)
    
    # Consider "heavy" if more than 10 LaTeX fragments
    return total_matches > 10


class Preprocessor(BasePreprocessor):
    name = "analysis_code_agent"  # MUST BE EXACTLY THIS - DO NOT MODIFY
    description = "Full-text chunk (with LaTeX stripped for math-heavy docs) + opening chunk + title-repetition chunk + top-3 density sentences chunk."

    OPENING_WORD_LIMIT = 200
    TITLE_REPEAT_COUNT = 5
    TOP_N_SENTENCES = 3

    def preprocess(self, docs: List[Document]) -> List[Chunk]:
        chunks = []
        for doc in docs:
            doc_id = doc.doc_id
            text = doc.text

            # Determine if document has heavy LaTeX pollution
            latex_heavy = has_heavy_latex(text)
            
            # Chunk 0: full document text
            # For LaTeX-heavy documents, strip the math notation to reduce token noise
            # and improve BM25 length normalization + IDF signal for content words.
            # For normal documents, use the text as-is to preserve all signal.
            if latex_heavy:
                chunk0_text = strip_latex(text)
            else:
                chunk0_text = text
            
            chunks.append(Chunk(
                chunk_id=f"{doc_id}_0",
                doc_id=doc_id,
                text=chunk0_text,
                metadata=doc.metadata,
            ))

            # Chunk 1: first ~200 words of the document (opening context)
            # Use cleaned text for LaTeX-heavy docs, raw text otherwise
            words = chunk0_text.split()
            if len(words) > self.OPENING_WORD_LIMIT:
                opening_text = " ".join(words[:self.OPENING_WORD_LIMIT])
                chunks.append(Chunk(
                    chunk_id=f"{doc_id}_1",
                    doc_id=doc_id,
                    text=opening_text,
                    metadata=doc.metadata,
                ))

            # Chunk 2: title repetition chunk
            title = extract_title(doc)
            if title:
                # Strip LaTeX from title too if needed
                if latex_heavy:
                    title = strip_latex(title)
                title_chunk_text = " ".join([title] * self.TITLE_REPEAT_COUNT)
                chunks.append(Chunk(
                    chunk_id=f"{doc_id}_2",
                    doc_id=doc_id,
                    text=title_chunk_text,
                    metadata=doc.metadata,
                ))

            # Chunk 3: top-N most information-dense sentences
            # Use the cleaned text for sentence extraction in LaTeX-heavy docs
            sentences = split_into_sentences(chunk0_text)
            if len(sentences) > self.TOP_N_SENTENCES:
                scored = [(sentence_density_score(s), i, s) for i, s in enumerate(sentences)]
                scored.sort(key=lambda x: (-x[0], x[1]))
                top_sentences = scored[:self.TOP_N_SENTENCES]
                top_sentences.sort(key=lambda x: x[1])
                density_text = " ".join(s for _, _, s in top_sentences)
                chunks.append(Chunk(
                    chunk_id=f"{doc_id}_3",
                    doc_id=doc_id,
                    text=density_text,
                    metadata=doc.metadata,
                ))

        return chunks
\end{lstlisting}

\begin{lstlisting}[style=python,caption={Preprocessing code from case study 2, generated for TipOfTongue split of CRUMB.},label={lst:case-study-code-2}]
import sys, pathlib
sys.path.insert(0, str(pathlib.Path(__file__).parents[2] / "evaluation"))
from typing import List
from schema import Document, Chunk
from base import BasePreprocessor

import re


def clean_wiki_markup(text: str) -> str:
    """
    Clean Wikipedia markup to extract plain text.
    """
    # Remove <ref>...</ref> blocks entirely (including multiline)
    text = re.sub(r'<ref[^>]*>.*?</ref>', ' ', text, flags=re.DOTALL | re.IGNORECASE)
    # Remove self-closing <ref ... />
    text = re.sub(r'<ref[^/]*/>', ' ', text, flags=re.IGNORECASE)

    # Remove [[Category:...]] and [[File:...]] and [[Image:...]] tags
    text = re.sub(r'\[\[(Category|File|Image|Media):[^\]]*\]\]', ' ', text, flags=re.IGNORECASE)

    # Convert [[link|display text]] -> display text
    text = re.sub(r'\[\[(?:[^\|\]]*\|)([^\]]+)\]\]', r'\1', text)

    # Convert [[link text]] -> link text
    text = re.sub(r'\[\[([^\]]+)\]\]', r'\1', text)

    # Handle nested templates iteratively
    prev = None
    while prev != text:
        prev = text

        def replace_template(m):
            inner = m.group(1)
            parts = inner.split('|')
            values = []
            for part in parts[1:]:
                part = part.strip()
                if '=' in part:
                    key, _, val = part.partition('=')
                    val = val.strip()
                    if val and not re.match(r'^https?://', val) and not re.match(r'^\d+$', val):
                        values.append(val)
                else:
                    if part and not re.match(r'^https?://', part):
                        values.append(part)
            return ' '.join(values) if values else ' '

        text = re.sub(r'\{\{([^{}]+)\}\}', replace_template, text)

    # Remove any remaining {{ or }} characters
    text = re.sub(r'\{\{|\}\}', ' ', text)

    # Remove section header markup (== Header ==, === Sub ===, etc.)
    text = re.sub(r'={2,}([^=]+)={2,}', r'\1', text)

    # Remove remaining HTML tags
    text = re.sub(r'<[^>]+>', ' ', text)

    # Remove wiki table markup
    text = re.sub(r'\{\|.*?\|\}', ' ', text, flags=re.DOTALL)
    text = re.sub(r'^\s*[|!].*$', ' ', text, flags=re.MULTILINE)

    # Remove bare URLs
    text = re.sub(r'https?://\S+', ' ', text)

    # Remove excessive punctuation artifacts from markup removal
    text = re.sub(r'[|]{2,}', ' ', text)
    text = re.sub(r'\[|\]', ' ', text)

    # Normalize whitespace
    text = re.sub(r'\s+', ' ', text).strip()

    return text


def extract_section(text: str, section_names: List[str]) -> str:
    """
    Extract a section from the cleaned text by looking for section header patterns.
    Returns the section content or empty string if not found.
    Works on the raw (pre-cleaned) text to capture section boundaries.
    """
    # Build a pattern that matches any of the section names
    names_pattern = '|'.join(re.escape(name) for name in section_names)
    # Match section header and capture content until next same-level or higher header
    pattern = re.compile(
        r'={2,}\s*(?:' + names_pattern + r')\s*={2,}\s*(.*?)(?=={2,}[^=]|$)',
        re.IGNORECASE | re.DOTALL
    )
    match = pattern.search(text)
    if match:
        return match.group(1).strip()
    return ''


def extract_sections_weighted(raw_text: str) -> str:
    """
    Build a weighted document:
    - Plot section: 3x repetition
    - Cast/Characters section: 2x repetition
    - Everything else: 1x
    
    All within a single chunk to preserve BM25 length normalization benefits.
    """
    # Extract plot section from raw markup
    plot_raw = extract_section(raw_text, ['Plot', 'Synopsis', 'Story', 'Plot summary'])
    # Extract cast/characters section from raw markup
    cast_raw = extract_section(raw_text, ['Cast', 'Characters', 'Cast and characters', 'Voice cast'])

    # Clean the extracted sections
    plot_clean = clean_wiki_markup(plot_raw) if plot_raw else ''
    cast_clean = clean_wiki_markup(cast_raw) if cast_raw else ''

    # Clean the full document
    full_clean = clean_wiki_markup(raw_text)

    # Build weighted text
    parts = []

    # Add full document once (baseline)
    parts.append(full_clean)

    # Add plot section 2 more times (total 3x since it's already in full_clean)
    if plot_clean:
        parts.append(plot_clean)
        parts.append(plot_clean)

    # Add cast section 1 more time (total 2x since it's already in full_clean)
    if cast_clean:
        parts.append(cast_clean)

    return ' '.join(parts)


class Preprocessor(BasePreprocessor):
    name = "section_weighted_repetition"
    description = (
        "Apply section-weighted text duplication within a single chunk: "
        "Plot section repeated 3x, Cast section repeated 2x, rest kept once. "
        "Uses in-chunk repetition to boost TF for plot/cast terms while "
        "preserving BM25 length-normalization benefits of a single long document."
    )

    def preprocess(self, docs: List[Document]) -> List[Chunk]:
        chunks = []
        for doc in docs:
            weighted_text = extract_sections_weighted(doc.text)

            chunks.append(
                Chunk(
                    chunk_id=f"{doc.doc_id}_0",
                    doc_id=doc.doc_id,
                    text=weighted_text,
                )
            )

        return chunks
\end{lstlisting}
\twocolumn

\end{document}